\documentclass[prb,twocolumn,superscriptaddress, showpacs]{revtex4}

\usepackage{amsmath}
\usepackage{amsfonts}

\usepackage{graphicx}
\usepackage{times}

\newcommand{\be}{\begin{equation} }
\newcommand{\ee}{\end{equation} }
\newcommand{\ba}{\begin{eqnarray} }
\newcommand{\ea}{\end{eqnarray} }

\begin{document}

\title{Quantum Disentangled Liquids}

\author{Tarun Grover}
\affiliation{Kavli Institute for Theoretical Physics, University of California, Santa Barbara, CA 93106, USA}
\author{Matthew P.\ A.\ Fisher}
\affiliation{Department of Physics, University of California, Santa Barbara, CA 93106}

\begin{abstract}
We propose and explore a new finite temperature phase of translationally invariant multi-component liquids which we call a ``Quantum Disentangled Liquid" (QDL) phase.  We contemplate  the possibility that in fluids consisting of
two (or more) species of indistinguishable quantum particles with a large mass ratio,
the light particles might ``localize" on the heavy particles. We give a precise, formal definition of this
Quantum Disentangled Liquid phase in terms of the finite energy density many-particle wavefunctions.  While the heavy particles are fully thermalized, 
for a typical fixed configuration of the heavy particles, the entanglement entropy of the light particles satisfies an area law; this implies that  the light particles have not thermalized.  Thus, in a  QDL phase, thermal equilibration is incomplete, and the canonical assumptions of statistical mechanics are not fully operative. We explore the possibility of QDL in water, with the light proton degrees of freedom becoming ``localized'' on the oxygen ions.  
We do not presently know whether a local, generic Hamiltonian can have eigenstates of the QDL form, and if it can not, then the non-thermal behavior discussed here will exist as an interesting crossover phenomena at time scale that diverges as the ratio of the mass of the heavy to the light species  also diverges
\end{abstract}

\pacs{05.30.-d, 03.65.Ud, 67.10.Fj}
\maketitle
\tableofcontents
\section{Introduction}

Conventional wisdom holds that in ordinary liquids at finite temperature, local thermal equilibration is established
on scales long compared to inelastic scattering lengths and times, and classical hydrodynamics is operative.  Quantum mechanics, which plays a role
on short scales and in setting the magnitudes of the crossover scales, can be safely ignored in describing 
qualitative behavior in the hydrodynamic regime.  For example, 
classical statistical mechanics should be sufficient in capturing the equilibrium and linear response behavior of ordinary room temperature liquids, e.g. water, on long length and time scales.

In this paper we question these assumptions. In particular,
we propose and explore a qualitative role that quantum effects might play in determining
the macroscopic behavior of finite temperature liquids.
We focus on two-component (or more general multicomponent) liquids in which there are two-species
of indistinguishable quantum particles with a large mass ratio, for example the ions and electrons that constitute the atoms in an atomic or molecular liquid (such as liquid Helium),
or two different ions with large mass ratio in a molecular fluid, such as oxygen and hydrogen ions in water.

To illustrate the general class of questions, consider the properties of Helium-3 fluid at temperatures above the superfluid transition (of several milliKelvin).  For temperatures below the Fermi energy (roughly $50\mathrm{K}$) He-3 is well described by Fermi liquid theory, while at higher temperatures a description as a non-degenerate -- essentially classical -- fluid would be appropriate.  At much higher temperatures, on the electron volt scale, as electrons are highly excited, significant ionization is expected, and a picture in terms of a plasma of electrons and He-3 nucleons would be appropriate.  At much higher temperatures still, the nucleons would start breaking into protons and neutrons, and with continued heating one should eventually enter into a regime described as a quark-gluon and electron plasma.  The canonical wisdom maintains that all such regimes are in the same phase, separated by crossovers rather than phase transitions \cite{footnote:firstorder}.  But might this conventional wisdom be incomplete or even incorrect?

In this paper we propose and explore a potentially new phase of finite temperature two-component quantum fluids in which the constituents have a large mass ratio.    A description of this new phase, which we will call a  ``Quantum Disentangled Liquid" (QDL), requires treating both species of particles quantum mechanically.
In the QDL phase  the light particles are, in a sense to be made more precise below,
``localized" \textit{on} the heavy particles, and due to quantum effects, do not thermalize. Due to the lack of full thermalization, a correct description requires a  ``microcanonical"
formulation wherein the fluid, confined to a very large box, is presumed to be in a pure quantum state,
perhaps an eigenstate, with a finite energy density above the ground state. 

In Section \ref{sec:singlecomp} we revisit the entanglement properties of single-component finite ``temperature" fluids
of indistinguishable quantum particles.   We focus on the nature of highly excited pure quantum states - 
with a finite energy density above the ground state. In a fully thermalized phase, for such states the entanglement entropy, for a spatial bi-partitioning,  will satisfy a volume law.  As we will argue, this high degree of spatial entanglement is encoded in the ``sign-structure" of these
finite energy-density states. Using this insight, we construct a  volume law wavefunction for bosons, which can be shown to possess a maximal entanglement entropy density  corresponding to an infinite ``temperature".

Section \ref{sec:multicomp} is devoted to a discussion of finite ``temperature" two-component quantum fluids,
with one species much more massive than the other.   We give a precise mathematical definition of the QDL phase which involves a diagnostic to be performed on the many-particle wavefunction.
Roughly, when the positions of the heavy particles are fixed, the light particles are in a ``disentangled"  area law phase.   For a lattice gas model of two boson species, we construct an explicit QDL wavefunction
which can be shown to satisfy all of the defining criterion of the QDL state.

The essence of the QDL phase can be conveyed by considering a wavefunction for a two-component
fluid of quantum particles which takes a Born-Oppenheimer \cite{born1927} type form.
To be explicit, consider a system of heavy, mass $M$, bosons and light, mass $m$, fermions,
with $m<< M$.    A Born-Oppenheimer wavefunction, expressed in terms of the coordinates of the
heavy and light particles, ${\bf R}_\alpha$ and ${\bf r}_\beta$, respectively,
takes the form,
\begin{equation}
\Psi(\{ {\bf R}_\alpha \}, \{ {\bf r}_\beta \} ) = \psi(\{ {\bf R}_\alpha \}) \det[ \Phi^R_{\beta} ( {\bf r}_{\beta^\prime}) ],
\label{eq:qdlBO}
\end{equation}
where the second term is a Slater determinant for the light fermions constructed from
single fermion orbitals, denoted, $\Phi^R_{\beta}({\bf r})$, which are  eigenstates of the effective Born-Oppenheimer Hamiltonian,
\begin{equation}
H = \sum_\beta [ \frac{ {\bf p}^2_\beta }{2 m} + V_R({\bf r}_\beta ) ],
\end{equation}
with effective potential,
\begin{equation}
V_R({\bf x}) = \sum_\alpha U^{Mm}({\bf x} - {\bf R}_\alpha)  .
\end{equation}
Here $U^{Mm}$ is the interaction potential between the light and heavy particles
and the heavy particle coordinates $\{ {\bf R}_\alpha \}$ are c-numbers. Above, the individual orbitals $\Phi^R_{\beta}$ have already been symmetrized in the coordinates $\{\bf{R}_\alpha\}$, so that $\Psi(\{ {\bf R}_\alpha \}, \{ {\bf r}_\beta \} ) $ is a valid wavefunction for bosons in the coordinates $\{\bf{R}_\alpha\}$, and for fermions  in the coordinates $\{\bf{r}_\beta\}$.

A caricature wavefunction for the conjectured QDL phase 
can be constructed by taking the first piece of the wavefunction in Eq.\ref{eq:qdlBO}, $\psi(\{ {\bf R}_\alpha \} )$,
to correspond to a finite-energy density fluid phase.  For the heavy bosons, a typical finite energy
density state for an isotropic liquid can be constructed from a permanent of plane waves, $\psi \sim \textrm{per}[e^{i {\bf k}_{\alpha} \cdot {\bf R}_{\alpha^\prime} } ]$, with wavevectors chosen with random directions, $\hat{\bf k}_\alpha$,
and with random amplitudes taken from some appropriate distribution,
for example a Boltzmann distribution,  $p_k \sim exp(-\epsilon_k/\overline{\epsilon})$ with $\epsilon_k = \hbar^2 k^2/2M$ and $\overline{\epsilon}$ corresponding to an ``effective temperature".
A simple wavefunction which describes the proposed QDL phase of two-component quantum particles
should thus take the form,
\begin{equation}
\Psi_{QDL}(\{ {\bf R}_\alpha \}, \{ {\bf r}_\beta \})  = \hskip0.1cm  \textrm{per}[e^{i {\bf k}_{\alpha} \cdot {\bf R}_{\alpha^\prime} } ]  \det[ \Phi^R_{\beta} ( {\bf r}_{\beta^\prime})  ] ,
\end{equation}
Due to the random wavevectors, the typical positions of the heavy particles will themselves be in an essentially random configuration, appropriate to a finite energy-density fluid phase.
As a result, the effective potential $V_R({\bf x})$ in the Born-Oppenheimer Hamiltonian for the light fermions will be an essentially random potential.  And one expects that for strong inter-species interaction
energy $U^{Mm}$ the low energy single particle eigenstates will be spatially localized \cite{anderson1958}.
The Slater determinant wavefunction thus describes a localized phase of fermions.   
One can imagine generalizing the above Born-Oppenheimer wavefunction to incorporate
(weak) inter-fermion interactions, replacing the Slater determinant by an
interacting wavefunction for a  ``Many-Body-Localized'' (MBL) phase of fermions, as introduced in Refs. \cite{basko2006, gornyi2005}.  
In either case, under a spatial bi-partitioning, into regions $A$ and $B$ with linear dimension $L$, this localized eigenstates of fermions will
have an area law entanglement entropy, $S_A \sim L^{d-1}$.

In contrast, the permanent wavefunction for the heavy particles $\psi \sim \textrm{per}[e^{i {\bf k}_{\alpha} \cdot {\bf R}_{\alpha^\prime} } ]$, being built from (random) plane waves is expected to have a volume law entanglement entropy, $S \sim L^{d}$, due to the essentially random sign structure upon
varying the positions of the particles
(as we shall discuss in considerable detail in Section \ref{sec:singlecomp}).   For a given fixed typical configuration of the
coordinates of the light fermions, say $\{ \tilde{\bf r}_\beta \}$, the Slater determinant will have a magnitude and sign
that varies rapidly when the positions of the heavy particles are changed.
But since the Slater determinant is in any event multiplied by a heavy particle wavefunction, 
$\psi(\{ {\bf R}_\alpha \})$, with an essentially random sign structure, the full QDL wavefunction $\Psi_{QDL}(\{ {\bf R}_\alpha \} , \{ \tilde{\bf r}_\beta \})$,
is still expected to have a volume law entanglement entropy.

These are the two primary properties of a QDL wavefunction  - 
when the heavy particles are fixed in a typical configuration the light particles are in an area law phase,
while, with the light particle coordinates fixed, the heavy particles have a volume law entanglement entropy.
In essence, the heavy particles have thermalized, but the light particles, ``localized" by the
heavy particles, have not.

Going beyond wavefunctions, the generic stability of the QDL phase in
 local Hamiltonians of interacting particles, requires, at the very least,
the generic stability of the MBL  \cite{basko2006, gornyi2005} state.
While an Anderson insulator of non-interacting fermions in a random potential
can exist at finite energy densities - provided that all the single particle states entering the Slater determinant are localized - establishing the stability of the MBL, a finite-energy density state of {\it interacting} electrons, is comparatively much more non-trivial and recent numerical progress \cite{huse2007, znidari2008, monthus2010, berkelbach2010, pal2011, gogolin2011, canovi2011, rigol2007, buccheri2011, cuevas2012, luca2013, serbyn2013_1, pollman2012, vosk2013, huse_ogn, serbyn2013_2, swingle, bela, bahri} has corroborated the original picture presented in Refs.\cite{basko2006, gornyi2005}.  Returning to QDL, even the  stability of the interacting MBL to arbitrary perturbations does not imply the stability of the QDL phase, which is much more tricky.
As illustrated in the Born-Oppenheimer discussion, if the heavy particle mass is taken to infinity, their positions will act as a quenched random potential for the light quantum particles which,
if in an area law MBL state at finite energy density, will not thermalize.  For very large but finite mass ratio, $M/m$, when the heavy particles become quantum dynamical and are assumed to thermalize,
there are two possibilities for the typical finite-energy density eigenstates of the two component quantum system:  (i) The light particles also thermalize, among themselves,
at the same ``temperature" as the heavy particles - this is the conventional implicit assumption
underlying the hypothesis of quantum statistical mechanics, (ii) The light particles do {\it not}
thermalize - this is the QDL phase.   While the {\it wavefunction}
of the QDL phase has a well defined and precise definition, as detailed in Section III,
it might be that it simply is not possible for a (generic)  local Hamiltonian to have eigenstates of the QDL form.    If this is the case, the $M/m \rightarrow \infty$ limit would be singular, and the non-thermal behavior
of the light particles could only occur at strictly $M = \infty$.   Presumably, in this case,
one could define a typical thermal equilibration time for the light particles, say starting from the MBL state, and this time would diverge as $M \rightarrow \infty$. On times scales short compared to this
crossover time the QDL physics would be operative, but would break down at longer times.
Whether or not  this equilibration time can remain infinite at finite $M$ - i.e. whether the QDL phase can exist as a ``true phase" - is not something that we can presently establish.

Section \ref{subsec:mblanalogy} will be devoted to a more general discussion of the physics of the QDL phase,
drawing further analogies with the Many-Body-Localized phase of quantum particles in a random potential.
In particular, the (linear response) conductivity in an MBL phase with a conserved particle number is expected to vanish, even at finite ``temperature" - the particles cannot propagate.  Analogously, in a QDL phase where the two species carry opposite ``electric" charge, one expects that the (linear response) electrical conductivity should vanish identically. We will also discuss hypothetical experiments in QDL which can detect lack of thermalization of the light particles.

In Section \ref{subsec:energetics} we will briefly discuss the prospects of establishing whether a QDL phase can exist in a local Hamiltonian for a two-fluid system, and the feasibility of verifying a QDL phase in numerics. In Section \ref{sec:physicalsys} we will explore the possibility that pure water might be in a QDL phase,
where the protons are ``localized" by the heavier oxygen ions.   
This possibility is informed by the remarkable phenomenology of pure water - its incredibly high resistivity, and the intricate hydrogen bonding which strongly correlates the motion of the oxygen ions with the protons.

\section{Quantum Entanglement in finite ``temperature" fluids} \label{sec:singlecomp}

Consider a single component translationally invariant fluid of indistinguishable quantum particles
(fermions or bosons).   For now we will ignore any internal ``spin" quantum numbers of the particles.
We imagine that the dynamics of the fluid is controlled by an underlying quantum Hamiltonian.
The precise nature of the Hamiltonian will not be critical, but we can think of non-relativistic particles
moving in the continuum interacting with a translationally invariant (say, two-body) interaction, see below.   The quantum particles are assumed to be isolated from any thermal bath, for example
confined in a box.  Our primary focus will be on the nature of eigenstates,
with a finite energy-density above the ground state.   The eigenstate thermalization hypothesis \cite{deutsch1991, srednicki1994, tasaki, rigol} posits that a large sub-region (the ``system") will behave thermally - as if at finite temperature - if one traces
out the degrees of freedom outside this region (the ``environment").   We will review and discuss the
entanglement properties of pure states in such a system.

\subsection{Operator and wavefunction representations}

Since we will be discussing such finite temperature quantum fluids from several different angles,
it will prove convenient to introduce and define several different representations/models - employing first and second quantized operators and states/wavefunctions  in the continuum, as well as a lattice gas model.

\subsubsection{Second quantization in the continuum}

We will denote general particle field operators by, $\hat{d}_{\bf x}, \hat{d}^\dagger_{\bf x}$
which satisfy canonical (anti-)commutation relations, $[\hat{d}_{\bf x}, \hat{d}^\dagger_{{\bf x}^\prime}]_\pm = \delta({\bf x} - {\bf x}^\prime)$, with $\pm$ sign for fermions/bosons respectively.
When we wish to specialize to fermions or bosons we will replace
the operators, ${\hat d}_{\bf x}$ by the more familiar notations, ${\hat c}_{\bf x}$ for fermion operators and 
${\hat b}_{\bf x}$ for boson operators. 
In terms of the field operators a 
representative fluid Hamiltonian is given by,
\begin{equation}
\hat{H}  = \int_{\bf x} \hat{d}^\dagger_{\bf x} (  - \nabla_{\bf x}^2 /2 m ) \hat{d}_{\bf x}   + \int_{{\bf x},{\bf x}^\prime}U(| {\bf x} - {\bf x}^\prime |) {\hat n}_{\bf x}  {\hat n}_{\bf x} , \label{eq:Hsinglecomp}
\end{equation}
where we have defined a particle density operator, ${\hat n}_{\bf x} = {\hat d}^\dagger_{\bf x} {\hat d}_{\bf x}$.
Here the particles have mass $m$, and interact via a two-body interaction $U(|{\bf x}|)$.
While we are primarily interested in three-dimensional fluids, we do not need to specify the dimensionality here.

For bosons the ground state of this model will likely be either a boson superfluid or a boson crystal.
And the fermion ground state could be a Fermi liquid, a paired superfluid or a fermion crystal.
But we are {\it not} interested in ground states.  We will focus on highly excited states at 
energy densities well above any condensation or crystallization temperatures.   

As mentioned above, we employ a microcanonical ensemble in which the fluid
is confined within a box of volume $V$, possibly with periodic boundary conditions,
and isolated from any thermal bath.
We are interested in eigenstates, with a finite energy density above the ground state.   Since our model is time reversal invariant we can take the exact eigenstates to be real.

\subsubsection{Particle wavefunctions in the continuum}

To discuss first quantized wavefunctions expressed in terms of particle coordinates,
it is convenient to define
a complete set of orthogonal position kets,
$|  \{ {\bf r}_\alpha \} \rangle = \prod_{i=1}^{\mathcal{N}_m} \hat{d}^\dagger_{{\bf r}_i}|\textrm{vac}\rangle $ where $\alpha = 1,2,..., {\cal N}_m$ labels the particles, with ${\cal N}_m$ denoting the total number of (mass $m$) particles.
For fermions, these kets are normalized,
while the normalization for the boson states
depends on how many bosons occupy the same spatial point.  To dispense with these 
unimportant yet distracting subtleties, we assume the bosons have a hard core, with radius $a$ 
setting the size of the particles.
Associated with a general state vector $| \Psi \rangle$,
is a position space wavefunction $\psi(\{ {\bf r}_\alpha \}) = \langle \{  {\bf r}_\alpha \} | \Psi \rangle$ which is  symmetric/antisymmetric under exchange of any two particles, for bosons/fermions.

\subsubsection{Quantum lattice gas}

At times below it will be convenient to place the particles on a lattice.   
We define a lattice gas model in terms of creation/annihilation operators $\hat{d}^\dagger_j, \hat{d}_j$, where the subscript $j=1,2,...,{\cal N}$ labels the lattice sites.
We assume the lattice Hamiltonian has a  ``canonical" hopping term and local
interactions, 
\be
\hat{H} = - \sum_{ij} t_{ij} ( \hat{d}^\dagger_i \hat{d}_j + h.c.) + \sum_{ij} U_{ij} \hat{n}_i \hat{n}_j,
\ee
 where both $t_{ij} \ge 0$ and $U_{ij}$ are local.
Here, $\hat{n}_j = \hat{d}^\dagger_j \hat{d}_j$ is the on-site density operator  - 
the hard core condition implies $\hat{n}_j$ has eigenvalues, $n_j=0,1$.
The model is taken to
respect the point and space group symmetries of the lattice, especially translations.

It is helpful to introduce position-space occupation number basis states,
$|\{ n_j \}  \rangle   = \prod_{j= 1}^{{\cal N}} [ \hat{d}^\dagger_{j} ]^{n_{j} } | \textrm{vac} \rangle  $,where  $n_j = 0,1$ denotes the absence/presence of a particle on site $j$.
For fermions one must choose some ordering for labeling the sites of the lattice - we do not
specify this unless it is necessary.
Associated with a general state vector  $| \Psi \rangle$
we can define a wavefunction in the occupation number basis, 
$\psi(\{ n_j\}) ) = \langle \{ n_j \} | \Psi \rangle$.

\subsection{Entanglement entropy for spatial bipartitioning}

Throughout we will assume the system is in a pure quantum state with density matrix, $\hat{\rho} = | \Psi \rangle \langle \Psi |$.   The density matrix can be either expressed in terms of the particle-position basis wavefunctions,
\begin{equation}
\rho(\{ {\bf r}_\alpha \} , \{ {\bf r}_\alpha^\prime \})  = \psi( \{ {\bf r}_\alpha \} ) \psi^*(\{ {\bf r}_\alpha^\prime \})  ,
\end{equation}
or the particle-occupation basis wavefunctions,
\begin{equation}
\rho(\{ n_j \} , \{ n_j^\prime \}) = \psi(\{ n_j \}) \psi^*(\{ n_j^\prime \}) .
\end{equation}

Consider a simple bi-partitioning into two spatial regions, denoted $A$ and $B$ respectively.
In the continuum we denote the volume of region $A$ as $V_A = L_A^d$, the ``system", and of region  $B$ as ${V}_B$, the ``environment", with ${V}_A + {V}_B = {V}$.  
On the lattice ${\cal N}_A$ and ${\cal N}_B$ denote the number of sites in the two regions, with 
${\cal N}_A + {\cal N}_B = {\cal N}$.

The reduced density matrix for sub-region $A$ is obtained as usual by tracing out the particles in region $B$, 
\begin{equation}
\hat{\rho}_A = Tr_{B} [\hat{\rho} ].
\end{equation}
And the (Renyi) entanglement entropies follow,
\begin{equation}
S_{A,{\alpha}} = \frac{1}{1- \alpha} \ln Tr_A [ ( \hat{\rho}_A )^\alpha]  ,
\end{equation}
with $S_A^{vN} = \lim_{\alpha \rightarrow 1} S_{A,\alpha}$ the von Neumann entropy.

\subsection{Area and volume laws}

Numerous studies have shown that the {\it ground state} of boson systems described by 
the representative Hamiltonians above will have area law entanglement entropy \cite{bombelli, srednicki, eisert2008},
$S_A^{(0)} \sim L_A^{d-1}$.   Indeed, both the boson superfluid and boson crystal
ground states exhibit an area law, and this is believed to be a general property of (unfrustrated)
boson ground states.
A paired fermion superfluid or fermion crystal ground state will also have an area law, while the entanglement
entropy of a Fermi liquid ground state has a multiplicative log correction to the area law \cite{wolf2006, gioev2006}, $S_A^{FL} \sim L_A^{d-1} \ln L_A$.

The entanglement entropy for eigenstates with a finite energy density above the ground state are expected
to be qualitatively different.  These eigenstates, which offer a microcanonical description of a finite ``temperature" fluid, are generally assumed to thermalize 
(although see Section III, below).   
Indeed, the eigenstate thermalization hypothesis \cite{deutsch1991, srednicki1994, tasaki, rigol} maintains that when $V_A << V_B$, the reduced density matrix $\hat{\rho}_A$ is the same as a {\it thermal} density matrix for the sub-system $A$ at a temperature for which the thermodynamic internal energy equals the energy per unit volume of the pure state $| \Psi \rangle$.  That is, the large sub-system 
$B$ can be viewed as a thermal bath which effectively causes thermalization of the
much smaller ``system" in volume $A$.  In particular, equilibrium statistical mechanics employing
the reduced density matrix, ${\hat \rho}_A$ 
can be used to compute the various thermodynamic properties of the sub-system $A$.

A consequence of the eigenstate thermalization hypothesis is that the 
von Neumann entanglement entropy satisfies a volume law,
$S_A^{vN} = s {V}_A$, and that $s$ can be equated with the thermal entropy density in
the sub-system $A$.   Similarly, the Renyi entropies also satisfy a volume law scaling, $S_{A,\alpha} \sim V_A$.

The eigenstate thermalization hypothesis is tantamount to the assumption that statistical mechanics is valid in a large quantum system;  a pure quantum state has sub-systems which
are correctly described using conventional statistical mechanics in the canonical ensemble wherein the sub-system has thermalized with an appropriate heat bath at some temperature.
[In Section III we will introduce a new finite temperature phase of multi-component quantum particles
which only {\it partially} thermalizes.]

While the assumed volume law behavior for finite-energy density states is most plausible, it would be desirable to have explicit volume law wavefunctions in order to develop a physical picture for (quantum) thermalization.   In the next subsection we construct  
simple wavefunctions that exhibit area and volume laws.

\subsection{Wavefunctions for area and volume law entanglement}

For simplicity we focus now on boson wavefunctions.   For the lattice gas model, the most general state vector can be expressed as,
\begin{equation}
| \Psi \rangle = \sum_{ \{ n_j \} } \psi(\{ n_j \}) | \{ n_j \} \rangle .
\end{equation}
Since the wavefunction amplitudes can be taken as real,
they can be decomposed as,
\begin{equation}
\psi(\{ n_j \}) = | \psi(\{ n_j \}) |  sgn(\{ n_j \})  ,
\end{equation}
where $sgn(\{ n_j \}) = \pm 1$.

\subsubsection{Boson ground state wavefunctions: area law}

The simplicity of boson ground states, $\psi_0$,  (of the representative Hamiltonians above) is that their wavefunctions are nodeless, with  $sgn_0(\{ n_j \}) = +1$  for all $2^{\cal N}$ configurations.
The wavefunctions in terms of particle coordinates are likewise nodeless, $\psi_0( \{ {\bf r}_\alpha \}) \ge 0$ for any particle configuration.

The entanglement entropy of such non-negative boson ground states will generically follow the aforementioned area law.  To illustrate this, consider 
a state that is expressed an an equal weight superposition over all occupation-number basis states,
\begin{equation}
\psi_0(\{ n_j \}) = \frac{1}{\sqrt{2^{\cal N}}} .
\end{equation}
This state encodes large density fluctuations of the particles and should describe 
a boson superfluid.  While it has been established that boson superfluid states have an area rather than volume law entanglement \cite{barthel, max}, this can be explicitly verified for the above equal amplitude wavefunction,
which can be re-expressed as a direct product state,
\begin{equation}
\psi_0(\{ n_j \}) = \prod_{j=1}^{\cal N}  \frac{1}{\sqrt{2}} [ \delta_{n_j,0} + \delta_{n_j,1} ] .
\end{equation}
For this product state the entanglement entropy actually vanishes identically, $S_{A,\alpha} = 0$.
Slight modifications of this superfluid wavefunction would lead to small corrections giving
an area law \cite{barthel,max}, $S_{A,\alpha} \sim L_A^{d-1}$.

\subsubsection{Volume law wavefunctions}

We now wish to construct wavefunctions for bosons which exhibit a volume law entanglement entropy,
which are hopefully representative of typical finite energy density eigenstates in such systems.
While the ground state of the boson system is nodeless, excited states necessarily have
nodes, with positive and negative wavefunction amplitudes.
And very highly excited eigenstates  - with a finite energy {\it density} above the ground state  -  
should be riddled with nodes, likely exhibiting an intricate and rich sign structure.

Intuition for such sign structure intricacies can be gained by considering particle-coordinate eigen wavefunctions for the ideal Bose gas, which are expressed as permanents of plane wave states,
\begin{equation}
\psi_{IBG}( \{ {\bf r}_\alpha \}) = \textrm{per} [M]   ,
\end{equation}
with matrix,
\begin{equation}
M_{\alpha \beta} = \frac{1}{\sqrt{V}} e^{i {\bf k}_\alpha \cdot {\bf r}_\beta} ,
\end{equation}
with ${\bf k}_\alpha$ a set of wavevectors with $\alpha =1,2,...,{\cal N}$.  
While the ground state Bose condensate has all ${\bf k}_\alpha = {\bf 0}$,
one expects that a typical finite energy density eigenstate would correspond to choosing the
wavevectors ${\bf k}_\alpha$  to be essentially random, with directions $\hat{\bf k}$ equally weighted over
the unit sphere, and with amplitudes chosen from an appropriate distribution function, such as
$p_{k} \sim exp(-\epsilon_k/ \overline{\epsilon})$ with kinetic energy 
$\epsilon_{\bf k} = \hbar^2 k^2 /2m$, and $\overline{\epsilon}$ an average kinetic energy.
Such a finite energy wavefunction should correspond to an effective temperature
of order $k_B T_{eff} \sim \overline{\epsilon}$.
We note that the wavefunction $\psi_{IBG}$ can be made real by taking the wavevectors to come in
$\pm {\bf k}$ pairs.

It is most plausible that permanent wavefunctions, $\psi_{IBG}(\{ {\bf r}_\alpha \})$,  constructed from plane waves with random wavevectors,
should exhibit a very complicated, and essentially random sign structure, as the particle positions
are moved about.    

Based on this reasoning, one would anticipate that the finite energy density eigenstates for the Bose lattice gas, have an occupation-number wavefunction representation with essentially random signs.
To illustrate this we will explore a wavefunction for lattice bosons which consists of an equal amplitude superposition
of occupation basis states, but with randomly chosen signs:
\begin{equation}
 \psi(\{ n_j \})  = \frac{1}{\sqrt{2^{\cal N}}}  sgn(\{ n_j \})  \label{eq:signrandom},
\end{equation}
with the $sgn$ function chosen randomly $sgn(\{ n_j \}) = \pm 1$, independently for each
of the $2^{\cal N}$ configurations $\{ n_j =0,1\}$.

A short calculation (see Appendix \ref{sec:volumelaw}), shows that the second Renyi entropy $S_{A,2}$ corresponding to the wavefunction in Eq.\ref{eq:signrandom} for a spatial bipartitioning of the system into regions $A$ and $B$ is given by 

\begin{eqnarray}
S_{A,2} & = & -\log\left(2^{-{\cal N}_A} + 2^{-{\cal N}_B} - 2^{-({\cal N}_A + {\cal N}_B)} \right) \nonumber \\
& \approx & \ln(2)  \,{\cal N}_A.
\end{eqnarray}
where in the second line, we have assumed that  ${\cal N}_A << {\cal N}_B$ (``system" much smaller than ``environment"). This confirms that the maximally random-sign wavefunction exhibits a volume law
entanglement entropy for the von-Neumann entropy $S_A^{vN}$ as well (recall the inequality \cite{nielsen} $S_A^{vN} > S_{A,2}$).  Moreover, the {\it amplitude}, $\ln(2)$, is the maximum possible value.
When interpreted as a thermal entropy this corresponds to ``infinite temperature",
with all $2^{{\cal N}_A}$ states in region $A$ summed over with equal weight in the canonical
partition function.  It is natural that this maximally entropic state, corresponds
to the wavefunction with completely random and uncorrelated signs.  If one were to modify the wavefunction by introducing some (spatial) correlations into the sign structure (still with half the signs positive and half the signs negative), one would expect that
the coefficient of the volume law would be reduced \cite{footnote:bias}.

Before turning to the two-component quantum fluid, in Section III, where we propose, define and discuss
the QDL phase, it will be helpful to briefly review the entanglement entropy
in localized states of quantum particles moving in a spatially dependent quenched random potential.

\subsubsection{Area law with disorder: Many-Body Localized states}

While we are interested in translationally invariant fluids, it will be instructive to remind the reader of the properties of wavefunctions for bosons (or fermions) moving through a spatially random quenched potential, described, for example, by adding a term to the Hamiltonian of the form,
\begin{equation}
\hat{H}_{dis} = \sum_j V_j \hat{d}^\dagger_j \hat{d}_j .
\end{equation}
Here, for example, one might take $V_j$  to be a site-uncorrelated random potential.   When the disorder strength greatly exceeds the particles kinetic and interaction energies, the ground state wavefunction is expected to be localized, either in a fermion insulator \cite{anderson1958} or a ``Bose glass"
phase \cite{fisher1989}.   A caricature wavefunction for the localized fermion insulator ground state might be constructed in terms of a determinant
built from the lowest energy localized single particle orbitals, $\phi_\alpha({\bf r})$,
\begin{equation}
\psi_{MBL}(\{ {\bf r}_\alpha \})  =   \det [M] ,
\end{equation}
with the matrix $M_{\alpha \beta} = \phi_\alpha ({\bf r}_\beta)$.
While this wavefunction is an eigenstate for non-interacting fermions, it is generally
believed that the system will still retain a localized insulating ground state
with weak enough interactions.

While the ground state should be localized with strong enough disorder, 
recent work  \cite{basko2006, gornyi2005, huse2007, znidari2008, monthus2010, berkelbach2010, pal2011, gogolin2011, canovi2011, rigol2007, buccheri2011, cuevas2012, luca2013, serbyn2013_1, pollman2012, vosk2013, huse_ogn, serbyn2013_2, swingle, bela, bahri}  has suggested that such strongly disordered systems can have highly excited finite energy density states that are likewise
localized.   These so-called Many-Body-Localized (MBL) phases \cite{basko2006,gornyi2005}
while having a finite energy density above the ground state, as for a thermal state,  are
{\it not} thermalized.  Indeed,  the entanglement entropy for the eigenstates in
an MBL state is believed to satisfy an area law, $S_{A,\alpha}^{MBL} \sim L_A^{d-1}$.
Loosely speaking, the MBL states can be thought of as real-space direct product states of single particle localized orbitals.

As we shall see in the next Section, for certain finite ``temperature" translationally invariant multi-component fluids, a new kind of entanglement entropy can be defined, which does have an area law, despite the absence of any quenched disorder.

\section{Quantum Disentangled Liquid: A Possible new phase in two-component fluids?} \label{sec:multicomp}

We next consider a translationally invariant two-component fluid of indistinguishable quantum particles,
a set of heavy particles with large mass $M$ and a set of light particles with mass $m << M$.
Initially we will not specify the quantum statistics of the two sets of particles.
The Hamiltonian we have in mind is a generalization of Eq.\ref{eq:Hsinglecomp} consisting of kinetic energies
for the two particle types, along with short-range two-body interactions, 

\be 
\hat{H} = \hat{H}_0 + \hat{H}_{U} \label{eq:Hmulticomp}
\ee

with 

\begin{equation}
\hat{H}_0 = \int_{\bf x}    \hat{D}^\dagger_{\bf x} (  - \nabla_{\bf x}^2 /2 M ) \hat{D}_{\bf x}+  \int_{\bf x} \hat{d}^\dagger_{\bf x} (  - \nabla_{\bf x}^2 /2 m ) \hat{d}_{\bf x} ,
\end{equation}
\begin{equation}
\hat{H}_U =  \int_{{\bf x},{\bf x}^\prime}  [ U^{m}_{{\bf x} {\bf x}^\prime}  \hat{n}_{{\bf x}} \hat{n}_{{\bf x}^\prime}
+   U^{M}_{{\bf x} {\bf x}^\prime}  \hat{N}_{{\bf x}} \hat{N}_{{\bf x}^\prime} +
U^{mM}_{{\bf x} {\bf x}^\prime}  \hat{n}_{{\bf x}} \hat{N}_{{\bf x}^\prime}  ],
\end{equation}
where $\hat{n}_{\bf x} \equiv \hat{d}^\dagger_{\bf x} \hat{d}_{\bf x}$ and $\hat{N}_{\bf x} \equiv \hat{D}^\dagger_{\bf x} \hat{D}_{\bf x}$.

We will again focus on the microcanonical ensemble in which the fluids are
confined in a box with volume $V$, possibly with periodic boundary conditions,
and isolated from any thermal bath.
We will primarily be interested in the situation where the 
system is placed in an exact many-body eigenstate of the Hamiltonian with a finite energy density above the ground state.   Since our model is time reversal invariant we can take the exact eigenstates to be real.

It will again be convenient to consider first quantized particle wavefunctions,
which we can define via the position kets,
\begin{equation}
|  \{ {\bf R}_\alpha \},  \{ {\bf r}_\beta \} \rangle = \prod_{\alpha=1}^{{\cal N}_M} \hat{D}^\dagger_{{\bf R}_\alpha} 
\prod_{\beta=1}^{{\cal N}_m}   \hat{d}^\dagger_{{\bf r}_\beta}    | \textrm{vac} \rangle ,
\end{equation}
where ${\cal N}_M$ and ${\cal N}_m$ denote the number of heavy and light particles, respectively.    We make no assumptions about ${\cal N}_M,{\cal N}_m$ except that
the ratio ${\cal N}_M/{\cal N}_m$ remains finite in the thermodynamic limit.  
Denoting the state vectors as $| {\Psi} \rangle$
we have position space wavefunctions, $\Psi(\{ {\bf R}_\alpha \}, \{ {\bf r}_\beta \}) = \langle \{ {\bf R}_\alpha \}, \{  {\bf r}_\beta \} | \Psi \rangle$.
For bosons/fermions these wavefunctions will be appropriately symmetrized/antisymmetrized under exchange of any two particles.

As for the single component case it will prove convenient to also define a quantum lattice gas model,
in terms of operators ${\hat d}_j, {\hat D}_j$ with $j=1,2,.., {\cal N}$ labeling the lattice sites.
The analogous lattice tight binding Hamiltonian for the two-component fluid takes the form
${\hat H} = {\hat H}_0 + {\hat H}_U$ with kinetic energy,
\begin{equation}
\hat{H} = - \sum_{ij} ( t^M_{ij}  \hat{D}^\dagger_i \hat{D}_j + t^m_{ij}   \hat{d}^\dagger_i \hat{d}_j  + h.c. )
\end{equation}
where the positive hopping amplitudes satisfy the inequality, $t^m_{ij} >> t^M_{ij} >0$, and an interaction energy,
\begin{equation}
\hat{H}_U =  
\sum_{ij} [ U^{m}_{ij} \hat{n}_i \hat{n}_j + U^M_{ij}  \hat{N}_i \hat{N}_j 
+ U^{mM}_{ij} \hat{n}_i \hat{N}_j ],
\end{equation}
with the definitions $\hat{n}_j = \hat{d}^\dagger_j \hat{d}_j$ and 
$\hat{N}_j = \hat{D}^\dagger_j \hat{D}_j$.

We again introduce position-space occupation number basis states,
\begin{equation}
|\{ N_j \}, \{ n_j \}  \rangle   = \prod_{j=1}^{{\cal N}} [{\hat D}_j]^{N_j}  \prod_{j= 1}^{{\cal N}} [ \hat{d}^\dagger_{j} ]^{n_{j} } | \textrm{vac} \rangle  ,
\end{equation}
where  $N_j =0,1$ and $n_j = 0,1$ denotes the absence/presence of a heavy particle and light particle on site $j$, respectively.
Associated with a general state vector  $| \Psi \rangle$
we define a wavefunction in the occupation number basis, 
$\Psi(\{ N_j \},\{ n_j\}) ) = \langle \{ N_j \}, \{ n_j \} | \Psi \rangle$ .

\subsection{Definition of Quantum Disentangled Liquid} \label{subsec:defqdl}

We will assume that the system is in a translationally invariant pure state $\Psi(\{ {\bf R}_\alpha \}, \{ {\bf r}_\alpha \})$, a finite energy density eigenstate 
of the interacting Hamiltonian in Eqn.\ref{eq:Hmulticomp}.
We will assume that the 
wavefunction is real and has the  exchange properties appropriate for the particles' statistics.

We make the following assumptions about the wavefunction.
Under a spatial bi-partioning into region $A$ and $B$ with $V_A \ll V_B$, we will assume that the
entanglement entropy exhibits a volume law, that is for
\begin{equation}
\hat{\rho}_A = Tr_{R_B, r_B} | \Psi \rangle \langle \Psi |  ,
\end{equation}
the reduced density matrix varies as,
\begin{equation}
S_A = - Tr_{R_A,r_A} [ \hat{\rho}_A \ln \hat{\rho}_A ] \sim  V_A   .
\end{equation}

We will also assume that the light and heavy particles have a ``mutual" volume law entanglement,
under a partitioning between the two components.
Specifically, define a reduced density matrix by integrating out only the light particles throughout the whole
volume, $V=V_A + V_B$,
\begin{equation}
\hat{\rho}_M = Tr_{m} | \Psi \rangle \langle \Psi |  .
\end{equation}
Then define an entropy by tracing over the heavy particles, also throughout the whole volume of the system,
\begin{equation}
S_{Mm} = - Tr_{M} \hat{\rho}_M \ln \hat{\rho}_M .
\end{equation}
This ``inter-component" mutual entropy is assumed to vary with the whole volume of the system,
$S_{Mm} \sim V$.

\subsubsection{Spatial and inter-component bi-partioning diagnostic} \label{subsubsec:qdlentropy}

We now introduce the following diagnostic of such two-component wavefunctions.
It consists of several steps.  In step one, we construct the probability distribution for the positions 
of the heavy particles, independent of the position of the light particles,
\begin{equation}
{\cal P}( \{ {\bf R}_\alpha \} )  = \int \prod_{\beta=1}^{{\cal N}_m} d {\bf r}_\beta | \Psi( \{ {\bf R}_\alpha \} , \{ {\bf r}_\beta \} ) |^2  .  \label{eq:probheavy}
\end{equation}

Next, viewing ${\cal P}( \{ {\bf R}_\alpha \} )$ as a probability distribution, we draw sets of coordinates
$ \{ {\bf R_\alpha^{(1)}} \},\{ {\bf R_\alpha^{(2)}} \},  \{ {\bf R_\alpha^{(3)}} \},...$  and so on.
And for a given member of this ensemble we return to the original wavefunction and now imagine it as a wavefunction solely of the light particles, with the position of the heavy particles $\{ \tilde{\bf R}_\alpha \}$ playing the role of a fixed parameter. This requires re-normalizing the wavefunction as
\begin{equation}
| \Psi \rangle_{\tilde{R}} \equiv \frac{1}{\sqrt{ {\cal Z}_{\tilde{R}} }} | \Psi \rangle, \label{eq:wfnqdl}
\end{equation}
with 
\begin{equation}
{\cal Z}_{\tilde{R}} = \int \prod_{\beta=1}^{{\cal N}_m} d {\bf r}_\beta |  \Psi( \{ \tilde{\bf R}_\alpha^{(n}) \} , \{ {\bf r}_\beta \} ) |^2 .
\end{equation}
so that $\int \prod_{\beta=1}^{{\cal N}_m} d {\bf r}_\beta |  \Psi_{\tilde{R}}( \{ {\bf r}_\beta \} ) |^2  = 1$.

We then define a density matrix,
\begin{equation}
\hat{\rho}^{ \tilde{R}}  = | \Psi \rangle_{\tilde{R}} \, {}_{\tilde{R}}
\langle \Psi |  .
\end{equation}
and with the coordinates of the heavy particles fixed, construct a reduced matrix from $\hat{\rho}^{ \tilde{R}} $ by tracing out the light particles in region $B$,

\begin{equation}
\hat{\rho}_A^{\tilde{R}} = Tr_{{\bf r}_B}  \hat{\rho}^{\tilde{R}}  .
\end{equation}

Next, we use the above reduced density matrix to compute the von Neumann entanglement entropy,

\begin{equation}
S_A^{\tilde{R}} =  - Tr_{{\bf r}_A}[ \hat{\rho}_A^{\tilde{R}} \ln \hat{\rho}_A^{\tilde{R}} ] .
\end{equation}
Note that here we have performed a trace solely over the light particles within region $A$. The von Neumann entropy thus defined depends on the coordinates of all the heavy particles
(both in regions $A$ and $B$), which we are denoting
$\tilde{R} = \{ \tilde{\bf R}_\alpha \}$.  

Finally, we ensemble average $S_A^{\tilde{R}}$ over the probability distribution
function ${\cal P}( \{ {\bf R}_\alpha \} )$   of the heavy particles (Eqn.\ref{eq:probheavy}):

\begin{equation}
S_A^{m/M} = \int \prod_{\alpha=1}^{{\cal N}_M} d \tilde{\bf R}_\alpha  {\cal P}(\tilde{R})  S_A^{\tilde{R} }   .
\end{equation}
Here the superscript $m/M$ is to indicate that this entropy refers to the entanglement entropy obtained by  spatially bi-partitioning the light particles, with mass $m$ and coordinates ${\bf r}_\alpha$, given a fixed position of the heavy particles with mass $M$ and coordinates ${\bf R}_\alpha$, only afterwards averaging over the distribution of the heavy particle positions.

By simply switching the coordinates, ${\bf r} \leftrightarrow {\bf R}$,
we can carry through this procedure to obtain the entropy of spatial entanglement of the
heavy particles given we fixed the positions of the light particles, which we will denote
$S_A^{M/m}$.

In the  ``conventional" description of a translationally invariant two-component liquid phase, one would typically {\it assume}
that the light and heavy particles have both thermally equilibrated,
implying a volume law for both entropies,
\begin{equation}
S_A^{m/M} \sim L_A^d  ;  \hskip0.4cm  S_A^{M/m} \sim L_A^d  ;  \hskip0.5cm  (\textrm{Thermal}) .
\end{equation}
These volume law behaviors imply that each species of particle contributes a piece to the entropy per unit volume in region $A$.

We are finally in a position to give a precise definition of the Quantum Disentangled Liquid.
The QDL phase is described by finite-energy density wavefunctions each of which has a volume law under a full spatial bi-partioning,
$S_A \sim V_A$, an inter-component entanglement varying as the volume of the whole system,
$S_{Mm} \sim V$, while for each wavefunction the two spatial/inter-component  bi-partioning entropies behave \textit{differently} - fixing the light particles leaves the heavy particles in a spatial volume law, \textit{but with the heavy particles fixed the light particles are in a spatial area law},
\begin{equation}
S_A^{m/M} \sim L_A^{d-1}  ;  \hskip0.4cm  S_A^{M/m} \sim L_A^d  ;  \hskip0.5cm  (\textrm{QDL}) .
\end{equation}

\subsection{Illustrative wavefunction for the QDL } \label{subsec:qdlwfn}

In order to construct an explicit wavefunction that we can prove is in a QDL phase, it will be convenient to
focus on the lattice model with both heavy and light boson particles.
Consider the boson wavefunction in the occupation number basis,
\begin{equation}
\Psi_{QDL}(\{ N_j\}, \{ n_j \} ) = \psi(\{ N_j \} ) \prod_{j=1}^{\cal N} \frac{1}{\sqrt{2}} [ \delta_{n_j,0} + e^{i \pi N_j} \delta_{n_j, 1} ]. \label{eq:qdlwfn}
\end{equation}
Here, the first term on the right hand sign, $\psi(\{ N_j \})$,  is a wavefunction for the heavy particles alone, and we assume that this is in the ``random-sign" volume law state from Section \ref{sec:singlecomp},
$\psi(\{ N_j \})  = 2^{-{\cal N}/2}   sgn(\{ N_j \})$
with random signs, $sgn(\{ N_j \}) = \pm 1$, chosen independently for each
of the $2^{\cal N}$ configurations $\{ N_j =0,1\}$.

We must now check each of the conditions for the QDL state.
Below, when convenient, we will employ the simplified notation,
$\Psi_{QDL}(N,n)$.  

We first consider tracing out the light degrees of freedom throughout the whole system,
$\hat{\rho}_M = Tr_m | \Psi \rangle_{QDL}\, {}_{QDL} \langle \Psi |$,
writing
\begin{equation}
\rho_M(N,N^\prime) = \sum_{\{ n_j \}}  \Psi_{QDL}(N,n) \Psi^*_{QDL} (N^\prime, n)  ,
\end{equation}
and explicitly perform the summation over ${\{ n_j \}}$ to get,
\begin{equation}
\rho_M(N,N^\prime) = \psi(N) \psi^* (N^\prime) \prod_j \frac{1}{2} (1 + e^{i \pi (N_j + N_j^\prime)})  .
\end{equation}
To assess the inter-species entanglement entropy we consider,
\begin{equation}
Tr_M (\hat{\rho}_M^2) = \sum_{ \{ N_j \} \{ N_j^\prime \} } \rho_M(N,N^\prime) \rho_M(N^\prime,N) .
\end{equation}
and find,
\begin{equation}
Tr_M (\hat{\rho}_M)^2 = \sum_{\{ N \}} |\psi(N)|^4 = 2^{{\cal N}} 2^{-2 {\cal N}} = 2^{-{\cal N}} ,
\end{equation}
giving the required volume law for the inter-species (2nd Renyi) entanglement entropy,
$S_{Mm,2} = - \ln Tr_M (\hat{\rho}_M)^2 = (\ln 2) {\cal N}$.

Next we consider a spatial bipartition to check that the wavefunction $\Psi_{QDL}$ has
a volume law spatial entanglement, when both the heavy and light particles are traced out in a region $B$.  It is convenient to introduce the notation, $N_A = \{ N_{j \in A} \}$ and $N_B = \{ N_{j \in B} \}$,
and similarly for $N \rightarrow n$,
so that we can write the reduced density matrix, $\hat{\rho}_A = Tr_B [\hat{\rho}]$, in a shorthand notation as,
$\rho_A \equiv \rho_A(N_A, n_A, N_A^\prime, n_A^\prime)$ with,

\begin{equation}
\rho_A = \sum_{N_B,n_B} \Psi_{QDL} (N_A,N_B,n_A,n_B) \Psi^*_{QDL} (N^\prime_A,N_B,n^\prime_A,n_B) .
\end{equation} 
The summation over $\{ n_B \}$ can be performed explicitly to give,
\begin{equation}
\rho_A = f(N_A,n_A) f(N_A^\prime, n_A^\prime) \sum_{N_B} \psi(N_A,N_B) \psi^*(N_A^\prime,N_B ) ,
\end{equation}
where we have defined,
\begin{equation}
f(N_A,n_A) = \frac{1}{\sqrt{2}} \prod_{j \in A} (\delta_{n_A,0} + e^{i\pi N_A} \delta_
{n_A,1} )  .
\end{equation}

We now examine $Tr_A [\hat{\rho}_A^2]$,
\begin{equation}
 \sum_{N_A,n_A,N_A^\prime,n_A^\prime} \rho_A(N_A, n_A, N_A^\prime, n_A^\prime) \rho_A(N^\prime_A, n^\prime_A, N_A, n_A) .
\end{equation}
We can use the identity,
\begin{equation}
\sum_{n_A,n_A^\prime} [f(N_A,n_A) f(N_A^\prime,n_A^\prime) ]^2 = 2^{2{\cal N}_A} 4^{-{\cal N}_A} = 1,
\end{equation}
to obtain the expression,
\begin{equation}
Tr_A [\hat{\rho}_A^2] = \sum_{N_B,N_B^\prime} | \sum_{N_A} \psi(N_A,N_B) \psi^*(N_A,N_B^\prime) |^2  ,
\end{equation}
With the presumed form of the heavy particle wavefunction, as a sum over all real-space occupation number configurations with a random sign, we obtain, 
\begin{equation}
Tr_{A}[\hat{\rho}_A^2] = 2^{-{\cal N}_A} + 2^{-{\cal N}_B} - 2^{-({\cal N}_A + {\cal N}_B)}  .
\end{equation}
This gives the required volume law for the spatial bi-partition Renyi entanglement entropy,
$S_{A,2} = - \ln Tr_{A}[\hat{\rho}_A^2] = (\ln(2)) {\cal N}_A$ for ${\cal N}_A << {\cal N}_B$. Again, this implies that von Neumann entropy $S_A$ also follows the volume law.

We next turn to the wavefunction diagnostic which involves freezing the coordinates of one species
of particle, and performing a spatial bi-partitioning into regions $A$ and $B$ and integration over the other particle type in region $B$.
Consider first fixing the positions of the heavy particles, choosing a configuration of the occupation numbers, $\{ \tilde{N}_j \}$.  Once the heavy particles are fixed in $\Psi_{QDL}(\tilde{N}, n)$,
the wavefunction in terms of the light particles is a product law in space.  Thus, the spatial entanglement entropy for the light particles will actually vanish, so that $S_A^{m/M} = 0$.
This is morally an area law, $S_A^{m/M} \sim L_A^{d-1}$.

Next we must fix the positions of the light particles, choosing some configuration
for $\{ \tilde{n}_j \} = 0,1$.  Let $\tilde{j}$ denote the subset of sites with 
$\tilde{n}_{\tilde j} = 1$.   The QDL wavefunction can then be written as,
\begin{equation}
\Psi_{QDL}( \{ N_j \}, \{ \tilde{n}_j \})  = C(\tilde{n})  \psi(\{ N_j \}) \prod_{\tilde{j}} e^{i \pi N_{\tilde{j}}} ,
\end{equation}
where the coefficient $C(\tilde{n}) = \prod_{\tilde{j}} (1/\sqrt{2})$.
Following Section\ref{subsubsec:qdlentropy}, this  requires normalizing this wavefunction
over the heavy particles, which simply sets $C(\tilde{n})=1$ for all configurations of the light particles,
giving $\tilde{\Psi}_{QDL} (N_A,N_B, \tilde{n}) = \psi(N_A,N_B) \prod_{\tilde{j}} exp(i \pi N_{\tilde{j}})$.

We must now spatially bi-partition, and integrate out the heavy particles in region $B$,
$\hat{\rho}_A^{\tilde{n}} = Tr_{N_B} | \tilde{\Psi}_{QDL} \rangle \langle  \tilde{\Psi}_{QDL} |$,
\begin{eqnarray}
\rho_A^{\tilde{n}}(N_A,N_A^\prime) &= &\prod_{\tilde{j}_A \in A} e^{i \pi(N_{\tilde{j}_A} + N^\prime_{\tilde{j}_A} )} \nonumber  \times \\ 
& &\sum_{N_B} \psi(N_A,N_B) \psi^*(N_A^\prime,N_B)  .
\end{eqnarray}

Finally, we can compute the trace over the density matrix squared, giving,
\begin{equation}
Tr_{N_A} [(\hat{\rho}_A^{\tilde{n}})^2 ]   = 2^{-{\cal N}_A} + 2^{-{\cal N}_B} - 2^{-({\cal N}_A + {\cal N}_B)}  ,
\end{equation}
or a volume law for the second Renyi entropy,
\begin{equation}
S^{M/m}_{A,2} = - \sum_{\{ \tilde{n} \}} {\cal P}(\{ \tilde{n}_j \} ) \ln Tr_{N_A} [(\hat{\rho}_A^{ \tilde{n} })^2] = (\ln 2) {\cal N}_A  .
\end{equation}
Here ${\cal P}(\{ \tilde{n_j} \} )$ is the probability of finding the light particles in the configuration $\{ \tilde{n}_j \}$,
\begin{equation}
{\cal P}(\{ \tilde{n}_j \}) = \sum_{\{ N_j \} } | \Psi_{QDL}(\{ N_j \} ,\{ \tilde{n}_j \} ) |^2  .
\end{equation}
Again, this  implies volume law for the von Neumann entropy as well \cite{footnote:boundsvn}. Therefore, our illustrative wavefunction, Eqn.\ref{eq:qdlwfn}, indeed satisfies all the properties postulated for the Quantum Disentangled Liquid (QDL). 
\subsection{QDL wavefunction of the Born-Oppenheimer form}

We now consider a wavefunction for a QDL phase of two-component quantum particles moving in the continuum.  This wavefunction will be constructed by using a Born-Oppenheimer type approach \cite{born1927}.
To be concrete we take the heavy particles to be bosons and the light particles to be fermions.
We assume that the number of heavy and light particles, ${\cal N}_M,{\cal N}_m$, are comparable,
as are their mean interparticle spacings, $\ell_M^{-d} = {\cal N}_M/V$ and $\ell_m^{-d} = {\cal N}_m /V$.
We are interested in a wavefunction with average energy per particle above the ground state, $\overline{\epsilon}$, which satisfies, 
\begin{equation}
\frac{\hbar^2}{ M \ell_M^2} << \overline{\epsilon} << \frac{\hbar^2}{ m \ell_m^2}  .
\end{equation}
This corresponds to an ``effective temperature" where the heavy particles are in a ``classical"
regime while the light particles are in a ``quantum" regime - although, as throughout this paper,
we are always treating both heavy and light particles as quantum particles.
We will also assume that the heavy and light particle are interacting with one another via
a strong potential, $U^{Mm}({\bf x})$ with a characteristic energy scale large compared to the ``degeneracy"
energy scale for the light particles,
$U^{Mm}({\bf 0}) >> \hbar^2/m \ell_m^2$. 

Consider a wavefuction of the form,
\begin{equation}
\Psi^{BO}_{QDL}(\{ {\bf R}_\alpha \}, \{ {\bf r}_\beta \})  =  \hskip0.1cm \textrm{per}[e^{i {\bf k}_{\alpha} \cdot {\bf R}_{\alpha^\prime} } ]  \det[ \Phi^R_{\beta} ( {\bf r}_{\beta^\prime})  ], \label{eq:qdlBOwfn}
\end{equation}
where the indices $\alpha,\alpha^\prime$ run from  $1,2,...,{\cal N}_M$ while the indices $\beta,\beta^\prime$ run from $1,2,...,{\cal N}_m$.
The first term in this wavefunction denotes a permanent for the heavy bosons with wavevectors ${\bf k}_\alpha$ that are chosen with a random direction and
an amplitude pulled from an appropriate Boltzmann distribution, $p_k \sim exp(-\epsilon_k/\overline{\epsilon})$ with $\epsilon_k = \hbar^2 k^2/2M$.

The second term is a determinant wavefunction for the light fermion particles.
Here, the single particle orbitals, 
$\Phi^{R}_\beta({\bf r})$, denote the lowest energy single particle eigenstates for the light fermions
in the presence of a static configuration of the heavy particles, $\{ {\bf R}_\alpha \}$,
described by the single particle ``Born-Oppenheimer" Hamiltonian,
\begin{equation}
{\hat H}^{BO}_{c} = \int_{\bf x} {\hat c}^\dagger_{\bf x} ( - \nabla_{\bf x}^2/2m) {\hat c}_{\bf x} + \int_{\bf x} V_{\bf x} {\hat c}^\dagger_{\bf x} {\hat c}_{\bf x} ,
\end{equation}
where we have defined a potential energy,
\begin{equation}
V_{\bf x} = \sum_{\alpha = 1}^{{\cal N}_M}  U^{Mm}({\bf x} - {\bf R}_\alpha)  .
\end{equation}
It is important to emphasize that the coordinates of the heavy particles $\{ {\bf R}_j \}$ that
enter into the effective potential are c-numbers, {\it not} operators.

Since the heavy particles have been placed into a wavefunction appropriate for a fluid,
constructed from a permanent of randomly selected single boson plane waves, a typical configuration
of heavy particles that carries appreciable weight in the full wavefunction  will correspond to essentially  ``random" locations of the heavy particles.   Thus, in the effective Born-Oppenheimer Hamiltonian
for the light fermions, the effective potential $V_{\bf x}$ will vary randomly in space.
Under the assumption that the inter-component interaction energy scale is large compared to the
degeneracy energy of the light particles, $U^{Mm}({\bf 0}) >> \hbar^2 /m \ell_m^2$, all of the
low energy single fermion eigenstates in ${\hat H}_{c}^{BO}$ should be localized.

Due to this localization, for given fixed positions of the heavy particles, the Slater determinant
of the light fermions describes a localized phase.
While we have assumed that the light fermions are occupying the lowest energy localized eigenstates,
which is reasonable in the extreme limit, $\overline{\epsilon} << \hbar^2 /m \ell_m^2$,
this can be relaxed.  Provided the higher energy occupied orbitals in the fermion Slater determinant
are all spatially localized one expects that the properties of the QDL wavefunction should
not change qualitatively.  Even upon inclusion of interactions between the light fermions,
the many fermion wave function can still describe a localized phase - a  Many-Body-Localized phase \cite{basko2006,gornyi2005}.  

Since the MBL wavefunction of the light fermions has an area law entanglement entropy
under spatial bi-partitioning into regions $A$ and $B$, the entropy $S^{m/M}_A$ defined in Section IIIA 
obtained by fixing the positions of the heavy particles should satisfy an area law, $S^{m/M}_A \sim L_A^{d-1}$, one of the two important defining properties
of the QDL phase.    

On the other hand, consider fixing the positions of the light particles in the QDL wavefunction, with some typical configuration, denoted $\{ \tilde{r}_\beta \}$.  While the magnitude, and more importantly the sign, of the
Slater determinant $\det[\Phi^R_\beta (\tilde{\bf r}_{\beta^\prime})]$, for given fixed $\{ \tilde{r}_\beta \}$,
will vary appreciably as the coordinates of the heavy particles are moved about, since the
sign of the permanent term in Eq.\ref{eq:qdlBOwfn} has, in any case, an essentially random sign structure already,
one expects that the entanglement entropy for the heavy particles under spatial bi-partitioning 
should be a volume law, $S^{M/m}_A \sim L_A^d$.    Thus, the Born-Oppenheimer type wavefunction
defined above is indeed expected to describe a QDL phase.

\section{Physics of the QDL} \label{sec:QDL_Physics}

The key property of the QDL phase as defined above is that it is {\it not} fully thermalized.  Physically, if one is given the positions of the heavy particles, the light particles appear as if they were localized due to  the effective potential provided by the heavy particles. Therefore, the dynamics and the mutual entanglement of the light particles is largely dictated by the instantaneous positions of the heavy particles. 

\subsection{Analogy with Many Body Localized physics} \label{subsec:mblanalogy}

Consider the limit where the mass $M$ of the heavy particles is taken to infinity. In this limit,
the heavy particles can be taken as fixed ``impurities", and the problem reduces to
solving for the quantum dynamics of the light particles moving in the random background
potential of the infinitely heavy particles.   Let's assume that this background static random potential
is strong enough that the light particles are in a Many-Body-Localized (MBL) phase \cite{basko2006, gornyi2005} - that is, generally,
a finite energy density state within which the particles cannot move, and have zero conductivity (even at finite energy density).  In the MBL phase, the light particles are in area law.  In this phase equilibrium statistical
mechanics essentially breaks down - the entropy per unit volume is zero, even though the system
is at a finite energy density.  The microcanonical ensemble, when one can take the quantum system in  a pure state, is not equivalent to the canonical ensemble for a system in thermal equilibrium
with a bath.  Essentially, in the presence of the random potential 
the quantum spectrum of the light particles is locally discrete.  Due to this discreteness,
it is essentially impossible for energy to move about, and one part of the system
cannot act as a thermal reservoir for the remaining part of the system.

Due to the area law, one can think of the particles in the MBL phase as essentially being in a direct product state.  This state is a very low entangled state having an area law - indeed a generic 
finite energy density state will have a volume
law entanglement entropy.   It is important to emphasize that such ``disentangling"
requires quantum mechanics, and there is no classical analog to the MBL phase.

One can think of the QDL as a phase within which the light particles fail to thermalize due to the presence of the heavy particles. While the heavy particles can move around once their masses are finite, $M^{-1} \neq 0$, the light particles can move much more quickly,
and can become ``localized" by the (nearly) static potential of the heavy particles.
While this picture of localization is no longer strictly
valid once the heavy particles can move, the definition in the preceding section
of the QDL phase in terms of the wavefunction is still precise. Effectively, with the positions of the heavy particles ``fixed" in the many-particle wavefunction, the light particles are in an area law state. 

To expose the partial non-thermalization of the QDL, consider a hypothetical scenario where one suddenly turns off all interactions between the particles (intra-, as well as, inter-species). This allows one to assign a contribution to the entropy solely from the light particles, `$S_{light}$'. At the same time, since the heavy particles in the QDL \textit{are} thermalized, one can  also associate a well-defined temperature $T_{QDL}$ with the QDL, which can be measured prior to the moment interactions are turned off. This can be measured, for example, by connecting the QDL to a non-QDL ideal gas (``thermometer'') and measuring the temperature of this ideal gas when there is no transfer of energy between the two systems. Repeating the above process at different temperatures, one can obtain a $S_{light}(T_{QDL})$ curve. The main point is that since the light particles in the QDL are not thermalized, the function $S_{light}(T_{QDL})$ will, in general, be different for different ensembles of the system, and for a given ensemble, may even be independent of $T_{QDL}$, signaling a breakdown of the thermodynamics in QDL. Such a process might be realizable in a two-fluid cold atomic system with large mass ratio for the two components, for example,  a ${}^{133}$Cs-${}^{6}$Li mixture \cite{repp}. In such systems, the interactions can be tuned by sweeping the system through Feshbach resonances which changes the intra- and inter-species scattering length. Subsequently the entropy $S_{QDL}$ can be measured by adiabatically transforming the light particles into an ideal gas.

\section{Discussion and Open Issues}

\label{sec:discussion}

\subsection{Stability of the QDL phase?} \label{subsec:energetics}

Perhaps the most important question pertinent to our exploration is whether or not the QDL phase
can exist as a typical finite energy-density eigenstate in a generic, local two-component model Hamiltonian
with very heavy and very light quantum particles?
Straightforward (full) exact diagonalization is in principle possible - with the full set of exact eigenstates in hand the QDL wavefunction diagnostic could be implemented.
But this will clearly be very challenging, since ED will only be possible in very small systems
with very few particles - perhaps up to 12 sites in a 1d tight binding model of heavy and light fermions.
It is unclear how effective DMRG will be in exploring QDL physics in even simple toy models.
The DMRG is best suited in accessing low entanglement states, usually ground states,
while the QDL state has a volume law entanglement entropy.  Perhaps some variant of dynamical
DMRG could be helpful in using short-time dynamics to explore the expected small area law entanglement of the light particles, once the heavy particles are fixed.
It appears that numerical progress on the QDL will require modifications of existing algorithms/approaches, at the very least, and perhaps the development of new algorithms tailored
to access and evolve (typical) volume law wavefunctions.

\subsection{Exploring the possibility of QDL Physics in Water}  \label{sec:physicalsys}

It seems worthwhile to ask if a QDL phase existed in an actual fluid, what would be its experimental manifestations? To put our ideas into concrete form, below we ask could the pure water be in a QDL phase,  with the light proton degrees of freedom becoming ``localized'' on the oxygen ions? We then discuss the consequences if such was indeed the case, in light of some of the transport experiments on water.

The oxygen to proton mass ratio is $16$, so water offers, at the very least, a natural setting to explore possible QDL physics.
Moreover, while $16$ is not so tremendously large, the covalent oxygen-hydrogen bonds (together with the hydrogen bonds themselves) will strongly correlate the relative motion of the protons and the oxygen ions.
Strong correlations between the heavy and light particles are central to the QDL.
In addition, the proton must be treated quantum mechanically with a 
thermal de Broglie wavelength comparable to the inter oxygen-hydrogen separations.
One might be tempted to treat the heavier oxygen ions semiclassically,
using a Born-Oppenheimer approximation.  As discussed in the introduction, the Born-Oppenheimer potential produced by a random (fluid) configuration of the oxygen ions,
might ``localize'' the protons on the oxygen ions.  The corresponding localized Born-Oppenheimer wavefunction, would then take the QDL form in Eq (4).

Let's then imagine that water at the standard temperature and pressure (STP) is in a  QDL phase.  Consider perfectly pure water, with exactly twice as many hydrogen as oxygen ions, and with no other ``impurity" ions present.   The protons are the light, charge $e$, spin-1/2 fermions, and the oxygen ions are the heavy, charge $-2e$, spin-zero bosons. If in a QDL phase, water should be a ``perfect insulator" with a strictly {\it zero} (linear response)
electrical conductivity, despite being at finite temperature.
An applied electrical field which pushes the protons and oxygen ions in opposite directions, could not set up an electrical current (linear in the field) since the protons are ``enslaved"
by the oxygen ions.

Experimentally, salt water is a fairly good conductor \cite{hill}, with a conductivity
of order $10^{-1}$  S/cm (a $10$  $\Omega$-cm resistivity). The conduction is dominated by solvated ions (salts) and tap water
has a much smaller conductivity, typically $10^{-4}$ S/cm.  Data on ``pure" water, of interest to us here,
is hard to come by.  The often quoted  ``theoretical" value for the conductivity of {\it pure} water \cite{waterhandbook} is tiny,
$5.5 \times 10^{-8}$ S/cm, and measured conductivites in the $10^{-7}$ S/cm  ball park have been mentioned in the literature \cite{pashley}.   This conductivity is very, very small, some $13$ orders of magnitude
smaller than the conductivity of a metal.

It will clearly be challenging experimentally to ascertain the difference between 
$10^{-7}$ S/cm  and strictly zero conductivity, as in a QDL phase,
especially since even nominally ``pure" water contains unwanted solvated ions which
will conduct.  Some progress can be made by measuring the electrical conductivity of
``supercritical water" using high pressures\cite{shocked_water,temp_shocked_water}, up to $500$ kbar
using ``shock" methods.  Upon pressurizing (nominally) pure water the conductivity
is found to rise.  Remarkably\cite{shocked_water,temp_shocked_water}, for very high pressures (under ``shock") 
the conductivity increases dramatically by some $8$ orders of magnitude,
from $10^{-7}$ S/cm  at STP to $10^{-4}$ S/cm at $50$ kbar before saturating at $20$ S/cm above 200 kbar (estimated temperature of $5,000$ K).   At these temperatures/pressures, pure water conducts over $100$ times better than salt water!  If pure water at STP were in fact in an insulating QDL phase
with zero conductivity, at such pressures a highly conducting and presumably fully thermalized phase
has been reached.   
Another direction to explore is the effect of deuteration on the electrical conductivity of water. 
How well does heavy water conduct?   Further experiments on the electrical properties
of water could be most interesting.
\subsection{Relevance to Finite ``temperature" Quantum Field Theory?} \label{subsec:ftqft}
Viewed more broadly, a fluid of heavy and light quantum particles at finite ``temperatures" is a special case of a two-component finite ``temperature" Quantum Field Theory.
Of relevance here is a QFT Lagrangian with two quantum fields,
a ``light" field with strong quantum fluctuations interacting strongly
with a ``heavy" field which is in a ``semi-classical"
regime.

The eigenstate thermalization hypothesis (ETH) \cite{deutsch1991, srednicki1994, tasaki, rigol} 
posits that the reduced density matrix in a large sub-volume $A$ (the ``system") obtained by
tracing out the degrees of freedom in the even-larger sub-volume $B$ (the ``environment"),
equals the thermal-density matrix in region $A$;
\begin{equation}
\hat{\rho}_A  \equiv Tr_B |\psi \rangle \langle \psi |  = Z_A^{-1} e^{-\beta \hat{H}_{QFT}^A} .
\end{equation}
Here $ \hat{H}_{QFT}^A$ is the QFT Hamiltonian projected into region $A$,
$Z_A = Tr_A e^{-\beta \hat{H}_{QFT}^A}$ 
and $\beta$ is determined by the condition that the von Neumann entanglement entropy,
$S_{vN} = - Tr_A \hat{\rho}_A \ln(\hat{\rho}_A)$ satisfies a volume law, 
$S_{vN}  = s V_A$ with entropy density
$s$ equal to the thermal entropy density:
$s =  \partial_T (T \ln Z_A)/V_A$. 

Imagine a two-component QFT, with a strongly fluctuating quantum field coupled to
a ``semi-classical" quantum field,
with finite energy-density eigenstates
that are in a QDL phase.
Since the QDL wavefunction is not fully thermalized,  it {\it cannot}
be accessed using the canonical thermal-ensemble and the ETH is not fully operative. For a QFT in the QDL phase,
if one fixes the background ``semiclassical field",
the entanglement entropy of the strongly fluctuating quantum field satisfies an area law.
It might be especially  interesting to explore such ideas further in the context of an elementary quantum particle interacting with
a background semiclassical gravitational field.

\acknowledgments{We thank Boris Shraiman, Daniel Fisher, Frank Pollmann, Jim Garrison, Leon Balents, Matt Hastings, Max Metlitski,  Michael Fisher, Moshe Goldstein, Olexei Motrunich,  Ryan Mishmash, Steve Shenker, T. Senthil, Yvette Fisher for helpful discussions and especially David Huse for helpful discussions and critical reading of our manuscript.  This research was supported in part by the National Science Foundation under Grants DMR-1101912 (M.P.A.F.) and NSF PHY11-25915 (T.G.), and by the Caltech Institute of Quantum Information and Matter, an NSF Physics Frontiers Center with support of the Gordon and Betty Moore Foundation (M.P.A.F.)}


\begin{thebibliography}{x}
\bibitem{footnote:firstorder} There could, in principle, always exist first order transitions that  separate various regimes that are otherwise smoothly connected (for example, liquid-to-gas transition in a conventional fluid).
\bibitem{footnote:bias} Note also that if one were to ``bias" the signs so that the
positive signs occurred with some probability $p>1/2$ then one would find that $S_{A,2} \sim O(1)$.
\bibitem{footnote:boundsvn} The von Neumann entropy $S$ satisfies $S > S_2$ \cite{nielsen}. Therefore, $\sum_i p_i S_i > \sum_i p_i S_{2,i}$ for a set of positive numbers $\{p_i\}$.
\end{thebibliography}

\appendix

\section{Volume law  entanglement entropy for random-sign wavefunction} \label{sec:volumelaw}

The ``random-sign'' wavefunction is given by 

\begin{equation}
 \psi(\{ n_j \})  = \frac{1}{\sqrt{2^{\cal N}}}  sgn(\{ n_j \})  ,
\end{equation}
with the $sgn$ function chosen randomly $sgn(\{ n_j \}) = \pm 1$, independently for each
of the $2^{\cal N}$ configurations $\{ n_j =0,1\}$.

To demonstrate that this maximally random-sign wavefunction has a volume law entanglement
entropy, it will be convenient to introduce a short-hand notation,
\begin{equation}
\psi(n_A,n_B) \equiv \psi( \{ n_{j \in A} \}, \{ n_{j \in B} \})   .
\end{equation}
With this notation the density matrix is simply,
\begin{equation}
\rho(n_A,n_B,n_A^\prime,n_B^\prime)  =   \psi(n_A,n_B) \psi^*(n_A^\prime, n_B^\prime) .
\end{equation}
The reduced density matrix can be obtained by tracing over the sites in region $B$:
\begin{equation}
\rho_A (n_A,n_A^\prime)  = \sum_{n_B}  \rho(n_A,n_B,n_A^\prime,n_B)  ,
\end{equation}
where $ \sum_{n_B}$ denotes a summation over $n_j =0,1$ for  $j \in B$.

The volume law is easiest to establish from the second Renyi entropy, which requires computing,
\begin{equation}
Tr_{A} [ \hat{\rho}_A^2] = \sum_{n_A,n^\prime_A} \rho_A (n_A,n_A^\prime) \rho_A (n_A^\prime,n_A) .
\end{equation}
Re-expressing in terms of the wavefunctions gives,
\begin{equation}
Tr_{A}[\hat{\rho}_A^2] =  \sum_{n_A,n_A^\prime}  |   {\cal M}_A(n_A,n_A^\prime) |^2  = \sum_{n_B,n_B^\prime}  |   {\cal M}_B(n_B,n_B^\prime) |^2 ,
\end{equation}
with the definitions,
\begin{equation}
{\cal M}_A(n_A, n_A^\prime) = \sum_{n_B}   \psi(n_A,n_B)  \psi^*(n_A^\prime, n_B)   ,
\end{equation}
\begin{equation}
{\cal M}_B(n_B, n_B^\prime) = \sum_{n_A}   \psi(n_A,n_B)  \psi^*(n_A, n_B^\prime)   .
\end{equation}

For the random-sign wavefunction, ${\cal M}_B$ can be expressed as,
\begin{equation}
{\cal M}_B(n_B, n_B^\prime) = \frac{1}{2^{\cal N}} \sum_{n_A}   sgn(n_A,n_B)  sgn(n_A, n_B^\prime)   .
\end{equation}
Consider two cases, $n_B=n_B^\prime$ and $n_B \ne n_B^\prime$.
In the former case we have,
${\cal M}_B(n_B,n_B) = 2^{{\cal N}_A}/2^{\cal N}$.
For the latter case we can use the central limit theorem,
\begin{equation}
\langle |{\cal M}_B(n_B \ne n_B^\prime)|^2 \rangle  =  2^{{\cal N}_A}/2^{2{\cal N}} .
\end{equation}
Here, the brackets $\langle ... \rangle$ indicate an average over an ensemble of random-sign wavefunctions, but for a given typical random-sign wavefunction this form should hold in 
the large ${\cal N}_A$ limit.

Combining the above expressions gives,
\begin{equation}
Tr_{A}[\hat{\rho}_A^2] = 2^{-{\cal N}_A} + 2^{-{\cal N}_B} - 2^{-({\cal N}_A + {\cal N}_B)}  .
\end{equation}
which implies that the second Renyi entropy,  $S_{A,2} \approx \ln(2)  \,{\cal N}_A$, when ${\cal N}_A << {\cal N}_B$.

\end{document}